\documentclass[12pt]{iopart}
\usepackage{graphics}
\usepackage{graphicx}
\usepackage{amssymb}

\begin{document}

\title[Damage Spreading in a DLG Model]{Damage Spreading in a Driven Lattice Gas Model}

\author[M. L. Rubio Puzzo et.~al]{M. Leticia Rubio Puzzo$^1$, Gustavo P. Saracco$^1$, Ezequiel V. Albano$^2$$^,$$^3$}

\address{$^1$ Instituto de Investigaciones Fisicoqu\'{i}micas Te\'{o}ricas y
  Aplicadas (INIFTA), UNLP, CCT La Plata - CONICET, c.c. 16, Suc. 4, (1900) La
  Plata, Argentina.}
\ead{lrubio@inifta.unlp.edu.ar}
\address{$^2$ Instituto de F\'{i}sica de L\'{i}quidos y Sistemas Biol\'{o}gicos (IFLYSIB), UNLP, CCT La Plata - CONICET, calle 59 nro 789, (1900) La Plata, Argentina.}
\address{$^3$ Departamento de F\'{i}sica, Facultad de Ciencias Exactas, UNLP, La Plata, Argentina.}

\begin{abstract}

We studied damage spreading in a Driven Lattice Gas (DLG) model as a function of the temperature $T$, the  magnitude of the external driving field $E$, and the lattice size. 
The DLG model undergoes an order-disorder 
second-order phase transition at the critical temperature $T_c(E)$, such that the ordered phase is characterized by high-density strips running along the direction of the applied field; while in the disordered phase one has a lattice-gas-like behaviour. 
It is found that the damage always spreads for all the investigated temperatures and reaches a saturation value $D_{sat}$ that depends only on $T$. 
$D_{sat}$ increases for $T<T_c(E=\infty)$, decreases for $T>T_c(E=\infty)$ and is free of finite-size effects. 
This behaviour can be explained as due to the existence of interfaces between the high-density strips and the lattice-gas-like phase whose roughness depends on $T$. 
Also, we investigated damage spreading for a range of finite fields as a function of $T$, finding a behaviour similar to that of the case with $E=\infty$.

\end{abstract}

\submitto{Journal of Statistical Mechanics: Theory and Experiment}

\maketitle

\section{Introduction}
The statistical mechanics of equilibrium phenomena is very useful for understanding the thermodynamic 
properties of many-particle systems from a microscopical point of view. 
From its beginnings up to now, new developments and theories have 
enriched it, culminating in the renormalization-group approach \cite{Stanley,Cardy}.     
In nature, most many-particle systems are under far-from-equilibrium conditions,
and yet there is not a well-established theoretical framework to treat them, as in the case of their 
equilibrium counterpart.

In order to overcome this shortcoming, many attempts have been made to gain some insight into 
the far-from-equilibrium behaviour, e.g. by studying simple models that are capable
on capturing the essential non-equilibrium behaviour. Within this context, one of
the best known paradigms of far-from-equilibrium systems is the two-dimensional
driven lattice gas (DLG) model proposed by Katz, Lebowitz and Spohn
\cite{kls}. This model consists of a set of particles located in a
two-dimensional square lattice in contact with a thermal reservoir. Particles
exchange places with nearest-neighbour empty sites according to spin exchange,
i.e. the Kawasaki dynamics. Also, an external drive is applied, causing  the
system to exhibit non-equilibrium stationary states (NESS) in the limit of
large evolution times. 
If a half-filled two-dimensional system is considered (as in this
paper), and for low enough temperatures, the DLG model develops an ordered
phase characterized by strips of high particle density running along the driving
direction \cite{beate}. However, by increasing the temperature a second-order
non-equilibrium phase transition into a disordered (gas-like) phase takes
place. The critical temperature ($T_c$) depends on the value of the driving
field $E$, and in the limit of $E\rightarrow \infty$ one has $T_c\simeq 1.41$ $T_O$,
where $T_O$ is the Onsager critical temperature of the Ising model
\cite{ising}. 
The critical behaviour of the DLG model has been studied
by using many different techniques \cite{zia2010}, such as field theoretical
calculations \cite{gallegos,JS,lcardy,lcardy2}, Monte Carlo 
simulations \cite{ALGA,MAGA,ktl,wan,daquila12}, 
finite-size scaling methods \cite{ktl,wan}, and short-time dynamic scaling \cite{alsa,saracco09,reviewST}, 
but the complete understanding of this model is still lacking and has originated a long-standing controversy \cite{zia2010,gallegos,JS,lcardy,lcardy2,ALGA,MAGA,ktl,wan,daquila12,alsa,saracco09,reviewST}.
It should be noticed that Monte Carlo studies
of the DLG model are mostly focused on understanding the critical
behaviour of NESS for half-filled lattices.

From the theoretical point of view, it is interesting and challenging to study
the dynamic evolution of a very small perturbation in a non-equilibrium
system. 
One way to do this is to apply the concept of \textit{damage spreading}. 
Originally introduced by Kauffmann \cite{kauf69,kauf84}, this
method is based on the point-to-point comparison between two slightly different configurations of a
system that are allowed to evolve simultaneously. 
In order to achieve these configurations, one sample is initially
perturbed by slightly changing its configuration, so that it is called the ''damaged" sample,
while the original sample remains unperturbed. Then the time evolution of the
perturbation, defined as the difference between configurations, is
followed. In the long-time limit, the perturbation can either survive or
vanish, according to the values of the control parameters of the system. 
In some cases, it is known that the passage from the survival of the
perturbation to its vanishing is effectively an irreversible phase transition
that can be related to directed percolation processes \cite{broad,hinrichsen2000}. 

Damage spreading studies were originally applied to the Ising model, spin glasses
and cellular automata, but they have also been applied to magnetic systems
such as the Potts model with q-states, Heisenberg and XY models, two-dimensional trivalent
cellular structures, biological evolution, non-equilibrium models, opinion
dynamics, and small world networks (see e.g. \cite{rubio2008} and references therein, and for more recent results see \cite{reviewST,anjos09,rubio10a,rubio10b,rubio10c,bernard10,tomko10,lundow12}).

Within the broad context discussed above, the
goal of our work is to give an overall description of the damage spreading process
in the DLG model as a function of the control parameters, i.e. the
temperature and the field magnitude, and also of the lattice dimensions.  

The manuscript is organized as follows: the DLG model is
described in \Sref{model}, while in \Sref{damage} details of the damage spreading technique are
explained. The results are presented and discussed in \Sref{results}, and
finally our conclusions are stated in \Sref{conclusions}.

\section{The Model}
\label{model}

The DLG model \cite{kls} is defined on the square lattice of size
$L\times M$ with periodic boundary conditions 
along both directions. The driving field, $E$, is applied along the $M-$direction. 
Each lattice site can be empty or occupied by a particle. If
the coordinates of the site are ($i,j$), then the label (or occupation
number) of that site can be $\eta _{ij}=\{0,1\}$. The set of all occupation
numbers specifies a particular configuration of the lattice. 
The particles interact among them through a nearest-neighbour 
attraction with positive coupling constant
($J>0$). So, in the absence of any field, the Hamiltonian is given by
\begin{equation}
\label{eq(1)}
H=-4J \sum _{\langle ij;i'j'\rangle}\eta _{ij}\eta _{i'j'},
\end{equation}
where  $\langle . \rangle$ means that the summation is made over nearest-neighbour sites only. 

The attempt of a particle to jump to an empty nearest-neighbour site,
$W_{jump}$, is given by the Metropolis rate \cite{metro} modified by the
presence of the driving field, that is,
\begin{equation}
\label{eq(2)}
W_{jump}=\min[1,\, e^{[\Delta H-\epsilon _{1}E]/k_{B}T}],
\end{equation}
where $k_B$ is the Boltzmann constant, $T$ is the temperature of
the thermal bath, $\Delta H$ is the energy change after the particle-hole
exchange, and $\epsilon _{1}=(1,\, 0,\,-1)$ assumes these values when the
direction of the jump of the particle is against, orthogonal or along the driving
field $E$, respectively. 
The field is measured in units of $J$ and temperatures are given in units
of $J/k_{B}$. In this context, the critical temperature for the case with $E=\infty$ is $T_c\simeq 3.2$.
The dynamics imposed does not allow elimination of
particles, so the number of them is a conserved quantity. 
Also, in the absence of a driving field, the DLG model reduces to the Ising model with conserving
(i.e. Kawasaki) dynamics. 
For further details of the DLG model, see e.g. \cite{beate,zia2010,marrodick}.

\section{The Damage Spreading Method}
\label{damage}
The Damage Spreading (DS) method was initially introduced by Kauffman
\cite{kauf69,kauf84} to investigate the effects of tiny perturbations
introduced in the initial condition of physical systems on their final
stationary or equilibrium states. 
In order to implement the DS method in
computational simulations \cite{herr90,binderMC}, two configurations or
samples $S$ and $S'$, of a given stochastic model, are allowed to evolve
simultaneously. Initially, both samples differ only in the state of a small
number of sites. Then, the difference between $S$ and $S'$ can be considered
as a small initial perturbation or damage.
In order to give a quantitative measure of the evolution of the perturbation,
the ``Hamming'' distance or damage $D(t)$ is defined as
\begin{equation}
D(t)=\frac{1}{N}\sum_{i,j}^{N}1-\delta_{\eta_{ij}(t),\eta'_{ij}(t)}, 
\label{defdam}
\end{equation}
\noindent where $N=L\times M$ is the total number of sites in the lattices,
$\eta'_{ij}(t)$ ($\eta_{ij}(t)$) is the occupation number of site $(i,j)$ in
the sample $S'$($S$), and $\delta_{\eta_i(t),\eta'_i(t)}$ is the Kronecker
delta function. The sum runs over all sites in both samples, so $0\leq
D(t)\leq 1$.  
If the perturbation introduced in $S'$ is small ($D(t=0)
\sim O(1/N)$), there are two possible scenarios: (i) the perturbation
disappears after some time and $D(t\rightarrow \infty)\rightarrow 0$
in the thermodynamic limit;
or (ii) $D(t\rightarrow \infty)$ is finite and the perturbation is
relevant.
Thus, in some cases there may be a transition between a state where damage heals and a
state where the perturbation propagates throughout the system. Often, this is a continuous
and irreversible critical transition \cite{rubio2008} that is named the Damage Spreading transition. 

\noindent It is well known that the critical behaviour of the DS transition depends on the dynamic rules used to implement the algorithm (e.g. heat-bath, Glauber, Swendsen-Wang, Metropolis, and
Kawasaki dynamics), in Monte Carlo simulations \cite{hinrichsen2000}. 
This dependence can be explained in terms of the detailed balance condition
\cite{hinrichsen2000,binderMC} that assures that the system will arrive at
an equilibrium state, but it does not establish a unique way for the
dynamic evolution. 
For example, in the Ising model with Glauber or Metropolis dynamic rules, the
initial damage goes to zero below a damage temperature $T_D\approx T_O$, where
$T_O$ is the Onsager critical temperature, and it spreads above $T_D$, while
the opposite scenario ($D>0$ for $T<T_D$, and $D=0$ for $T>T_D$) is observed
when using the heat-bath dynamics.
It is worth mentioning that the Glauber, Metropolis, and heat-bath dynamics
are examples of non-conserved order parameter dynamic rules. 
On the other hand, a different behaviour is observed when conserved order
parameter rules, as e.g. in the case of the Kawasaki dynamics, are applied to
the Ising model.
In fact, by studying the two-dimensional Ising model, Glotzer and Jan
\cite{glotzer89} did not observe any phase transition in the damage spreading
probability at the critical temperature $T_O$, so they concluded that damage
always spreads. A few years later, Vojta \cite{vojta} studied the kinetic
Ising model with Kawasaki dynamics by using an effective-field theory and
Monte Carlo simulations. He found that two systems, whose initial
configurations differ only in a few sites, become completely uncorrelated in
the long-time limit. Moreover, he also found that the asymptotic average
damage is equal to $1/2$. 
To the best of our knowledge, the DS behaviour of the DLG 
has not been studied yet. 
However, in the broad context of the above-discussed results,  
it is expected that damage in the DLG will tend to the Ising
model with Kawasaki dynamics in the limit $E\rightarrow 0$, and therefore
$D(t\rightarrow \infty) \rightarrow 1/2$, but the role played by the driving field remains to be clarified and will provide valuable hints for the understanding of far-from equilibrium systems.

\section{Results and Discussion}
\label{results}

Monte Carlo simulations were performed by using square lattices of sizes $L \times
M$ with $30\leq L,$ $M \leq 480$ lattice units (LU) and by varying the aspect
ratio $L/M$. 
In order to introduce a perturbation into the system, NESS states
were generated after a very long time evolution of the system, typically
$t\geqslant 10^6$ Monte Carlo time steps (MCS), where an MCS is defined as
$L\times M$ attempts for a randomly chosen particle to jump into a neighbouring
site. Then, two samples of the system, $S$ and $S'$, are obtained. In one
of them, let us say $S'$, some particles are taken out and relocated in empty
positions, so that the initial configurations for each sample $S$ and $S'$ are
slightly different. After generating that damage, both systems are allowed to evolve with the
same sequence of random numbers. The total damage $D(t)$ given by equation \eref{defdam} is
measured as a function of time for different values of the external field
$0\leq E\leq 50$ and the temperature $T$
in the range $0.5 \leq T \leq 16$.

\begin{figure}[!tbp]
\begin{center}
\includegraphics[width=10cm,clip=true]{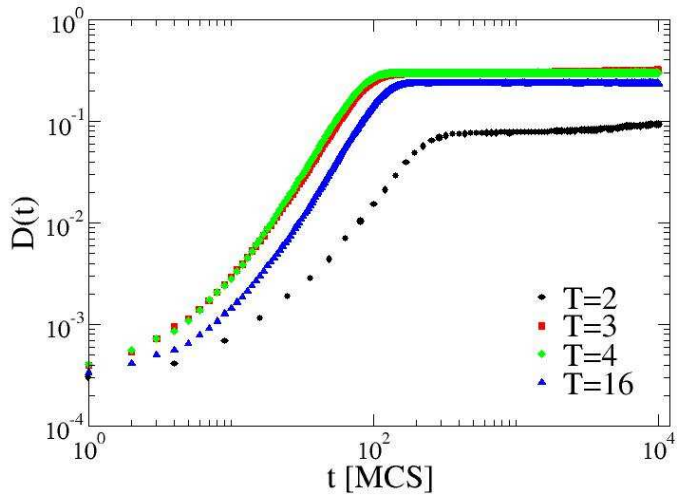}
\caption{Log-log plot of $D$ versus $t$, as obtained for a lattice of size $L \times M= 60\times 120$, and keeping $E=50 \approx \infty$ constant. The different values of the temperature $T$ are indicated in the legend.}
\label{fig1}
\end{center}
\end{figure}

\begin{figure}[!tbp]
\begin{center}
\includegraphics[width=10cm,clip=true]{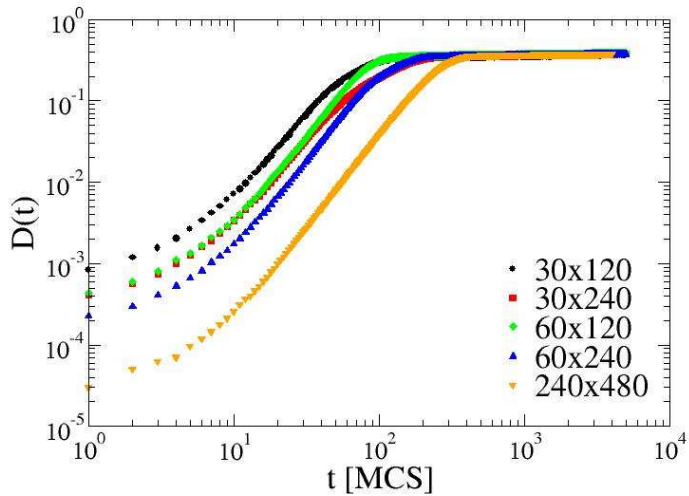}
\caption{Log-log plot of the temporal evolution of the damage obtained for $T=3.2\approx T_c (E=\infty)$, $E=50 \approx \infty$ and different lattice sizes $L \times M$, as indicated.}
\label{fig2}
\end{center}
\end{figure}

\Fref{fig1} shows the results of $D$ versus time obtained for a lattice of size $L \times M= 60\times 120$, $E=50 \approx \infty$ (notice that according to the transition rule given by equation \eref{eq(2)}, $E=50$ is in practice equivalent to $E\equiv \infty$) and different temperatures, as indicated. 
In all cases the initial damage is taken as $D(t=0)=1/LM$.
As can be observed, damage always spreads within the temperature range investigated and after some time it 
reaches a saturation value $D_{sat}$ that depends on the temperature. 
In order to establish the dependence of the saturation value $D_{sat}$ on the lattice size, we also performed simulations of the temporal evolution of damage for $E=50$, and $T=3.2\approx T_c(E=\infty)$, by taking different lattice sizes $L\times M$ and aspect ratios $L/M$. 
The results obtained are shown in \fref{fig2}. 
The absence of finite-size effects in the asymptotic or saturated value of $D$, $D_{sat}$, can clearly be observed and confirms that for $E=\infty$, $D_{sat}$ depends only on $T$, but it is no longer dependent on the lattice size and the aspect ratio.

\begin{figure}[!tbp]
\begin{center}
\includegraphics[width=10cm,clip=true]{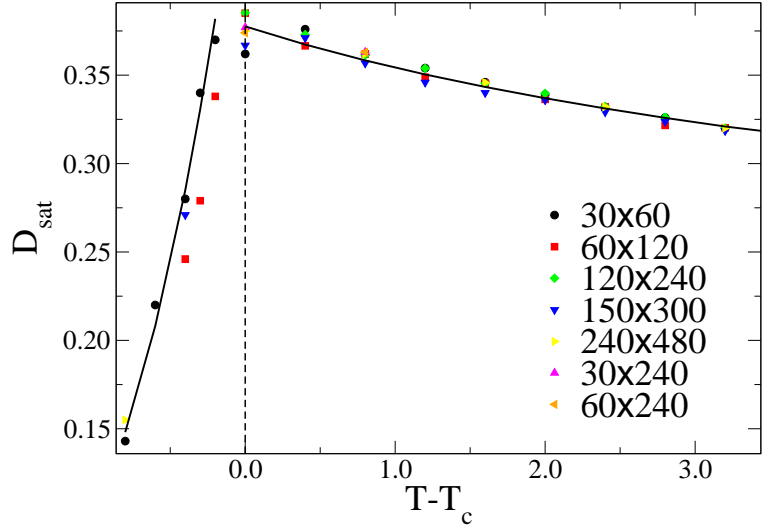}
\caption{Linear-linear plot of $D_{sat}$ versus the distance to the critical temperature $T-T_c$ obtained for $E=50 \approx \infty$ and different lattice sizes $L \times M$, as indicated.  The full lines are power-laws obtained fitting the data by means of a linear regression in each regime, while
 the vertical dashed line indicates the location of the critical temperature $T=3.2\approx T_c (E=\infty)$, in order to guide the eye.}
\label{fig3}
\end{center}
\end{figure}

\noindent By keeping $E=50$ and varying the temperature of the system, we then studied the
dependence of $D_{sat}$ on $T$ (see \fref{fig3}). 
For $T<T_c$, $D_{sat}$ increases
and close to $T\simeq T_c$ it shows a peak; subsequently it starts to decrease monotonically.  
It is important to notice that in both regimes, $D_{sat}$  behaves as a power law, i.e. $D_{sat} \sim \mid T-T_c\mid^{\kappa_{1,2}}$, where $\kappa_1=4.2(3)$ for $T<T_c$ and $\kappa_2=-0.23(2)$ in the $T>T_c$ regime. 
This increase and decrease behaviour can be explained in terms of the presence of solid strips with
interfaces between particles and empty sites. 
In fact, at low enough temperatures, the system is almost frozen in
the ordered phase, the high-density strips are quite compact, and
their interfaces are flat \cite{saralba}. For this scenario, the sample $S$ and its perturbed
counterpart $S'$ are almost identical, even in the long-time limit. 
So, the saturated damage, $D_{sat}$, is constrained to a small value. 
When the temperature is raised, the interface roughness increases \cite{saralba}, leading to the generation of
configurations where damage can spread. 
This process continues until $T\simeq T_c$, where the strips vanish in the disordered phase. In this way, $D_{sat}$ increases up to $T\simeq T_c$.
On the other hand, when $T\gtrsim T_c$ the system enters into a disordered
phase without well-defined strips, but large clusters of particles are still present, whose
interface sizes tend to zero in the limit $T\rightarrow \infty$.
Consequently, $D_{sat}$ also decreases in this limit due to the
lack of interfaces, as is shown in \fref{fig3} for $T>T_c$.
It is worth mentioning that similar results were obtained for external driving fields in the range 
$0.5\leq E \leq 10$ (not shown here for the sake of space).
\begin{figure}[!tbp]
\begin{center}
\includegraphics[width=10cm,clip=true,angle=0]{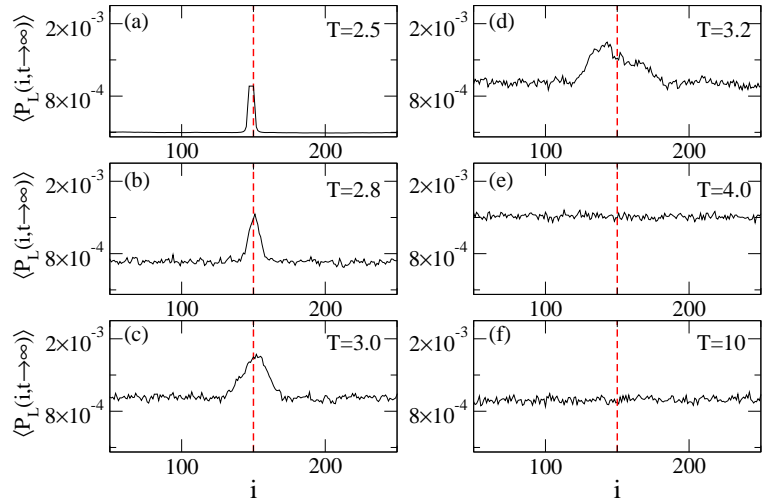}
\caption{Transversal damage profiles measured during the NESS for $t=10^7$
  MCS and obtained for the indicated temperatures. The lattice size is
  $L=300$, $M=150$. The dashed line indicates the initial position of the interface at $i=L/2$.}
\label{fig4}
\end{center}
\end{figure}
For a better illustration of the above statements, we recorded the damage profiles
in the transversal direction (i.e. perpendicular to the external driving field axis)
when the system had reached the NESS, which can be defined as
\begin{equation}
\langle P_L(i,t\rightarrow \infty)\rangle=\frac{1}{M} \sum_{j=1}^{M} 1-\delta_{\eta_{ij},\eta'_{ij}}  ,
\end{equation} 
\noindent where the averages $\langle.\rangle$ are taken over 100 independent
samples.
All simulations are started from the same initial configuration, where all particles are disposed in a
compact band on the left-hand side of the sample, i.e. for  $0<i<L/2$. The profiles were obtained by the following procedure: (i) a particle of the perturbed sample is removed from the interface at $i=L/2$ and placed at a randomly selected position in the remaining half of the lattice; (ii) the system is allowed to evolve for $t=10^7$ MCS up to NESS states, and (iii) the profiles $\langle P_L(i,t\rightarrow \infty)\rangle$ are registered for the next $t=10^7$ time steps. 
The obtained profiles are shown in \fref{fig4}. 
At low temperatures, for  $T<T_c$, (see panels (a) and (b)), the damage is concentrated in one peak that corresponds to the diffusion of damaged sites along the interface at $i=L/2$. 
Near and at the critical temperature $T\leq T_c$ (see panels (c)-(d)), the peak broadens. This behaviour can be attributed to the increase in the roughness of the interface with temperature (see below) \cite{saralba}. 
For  $T>T_c$ (panel (e)), the peak is missing, damage spreads and becomes spatially uniform. 
Finally, at $T\gg T_c$ (panel (f)), the damage decreases and also remains spatially uniform. 
It is worth mentioning that this is consistent with that shown in \fref{fig3}, and it is confirmed by the snapshots of the system (left panels) and the corresponding damaged sites (right panels), as shown in \fref{fig5}. In fact, by starting from a random configuration the system evolves to NESS. For $T<T_c$ we observe the formation of strips and the damage is naturally segregated at the interfaces between the high-density strips and the gas-like phase. Also, for $T>T_c$ the distribution of the particles in the system, as well as the damage, becomes more uniform.

\begin{figure}[!tbp]
\begin{center}
\includegraphics[width=10cm,clip=true]{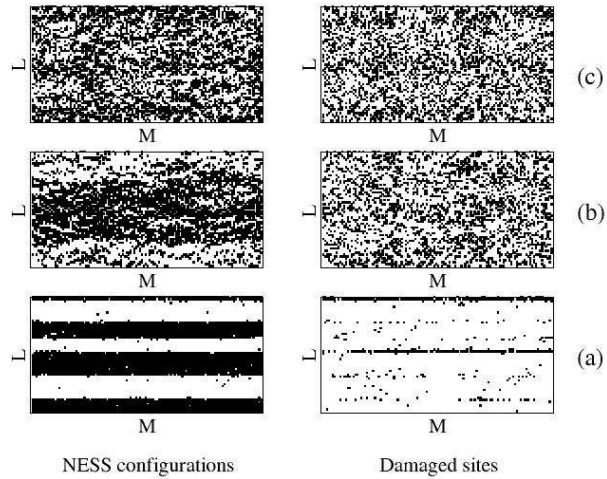}
\end{center}
\caption{\label{fig5}Snapshots of NESS (left-hand side panels) where the particles are
  represented by black dots. In the right-hand side panels, damaged sites are
  shown in black dots. Configurations corresponding to $E=50\approx \infty$, $L \times
  M= 60\times 120$, and obtained at (a) $T = 2<T_c$, (b) $T=3\approx T_c
  (E=\infty)$, and (c) $T = 4>T_c$.}
\end{figure}

\begin{figure}[!h]
\begin{center}
\includegraphics[width=10cm,clip=true,angle=0]{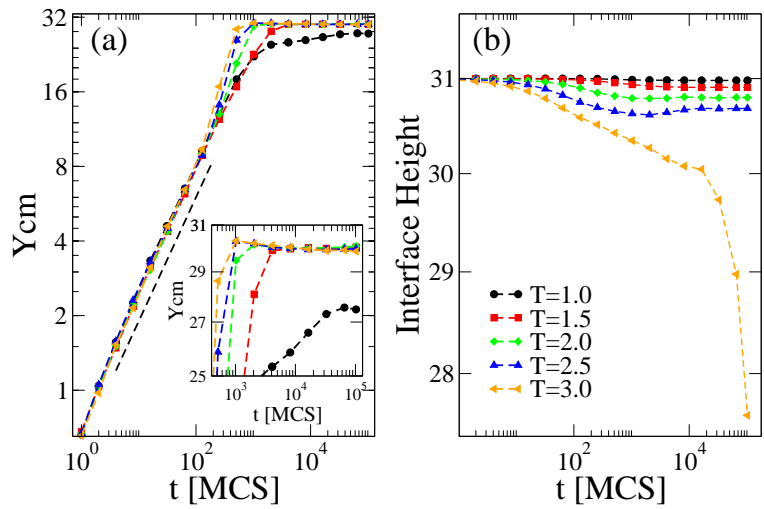}
\caption{(a) Time series of $Y_{cm}$ on a log-log scale. The dashed line has slope
1/2 and has been drawn for the sake of comparison. The inset shows the saturation regime of the data shown in the main panel. (b) Interface height as a function of time. In both cases the lattice size is $L\times M=120 \times 60$ and the temperatures employed are indicated.}
\label{fig6}
\end{center}
\end{figure}

Since the damage has a strong correlation with the interfaces of the system, it turns out to be reasonable that it can be related to its average position and roughness $W$, defined as the root-mean-square of the former \cite{saralba}. 
In order to study this relationship, we performed extensive simulations by starting the system with all the particles forming a compact strip (i.e. without holes inside) and placed at the centre of the sample, i.e. for $L/4\leq i \leq 3L/4$, but in this case  only one particle was removed from the middle,  so the initial damage is of order $O(1/LM)$ and it is placed at $i=L/2$, $1\leq j\leq M$. 
The time evolution of the coordinate of the damage centre of mass $Y_{cm}$ in the direction perpendicular to the driving was recorded for different temperatures in the range $1.0\leq T \leq 3.0$. \Fref{fig6} (a) shows the obtained results.  In the short-time regime, the damage behaves diffusively inside the strip, $Y_{cm}\approx t^{1/2}$ for all temperatures. Subsequently, a saturation value is reached at long times. \Fref{fig6} (b) exhibits the average position of the interfaces, which remains close to the centre of the sample. 
So, by comparing the average location of the damage (e.g. \fref{fig6} (a)) and the average position of the interface (e.g. \fref{fig6} (b)) one unambiguously concludes that the damage remains confined to the interface.
Furthermore, it is well known that in the DLG model the stationary value of the interface width $W$ grows with $T$ up to $T_c$  in the ordered phase \cite{saralba}, a fact that can also be observed in \fref{fig5}. 
This means that the interfacial configurations become rougher when $T$ approaches $T_c$ and are energetically favourable for the creation and spreading of damaged sites, explaining the growing regime shown in \fref{fig3} and \fref{fig4}, which can be confirmed by a direct inspection of both the left- and right- hand side panels of \fref{fig5}. Also, due to this fact one concludes that damage saturates when it approaches the interface position, even in the case of $T=3.0$ depicted in \fref{fig6} (b) (the error bars in this figure are not shown for the sake of clarity).

\begin{figure}[!tbp]
\begin{center}
\includegraphics[width=10cm,clip=true,angle=0]{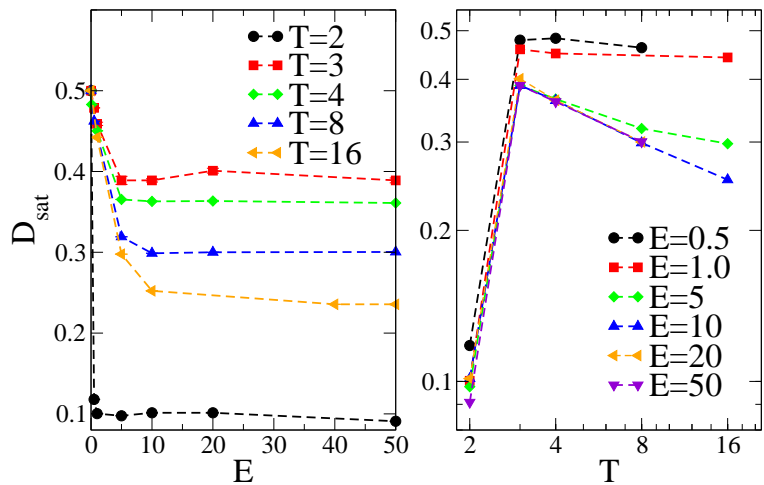}
\caption{Saturation damage $D_{sat}$ versus the external field $E$ (left-hand side panel, on a linear-linear scale) and the temperature $T$  (right-hand side panel, on a linear-log scale), obtained for different values of $T$ and $E$, indicated in the respective legends. The data are obtained by using  lattices of size $L \times M= 60\times 120$.}
\label{fig7}
\end{center}
\end{figure}

Finally, we performed further simulations at finite values of $E$ in
the range $0.25\leq E \leq 10$. \Fref{fig7} shows the dependence of $D_{sat}$ on the external
applied field $E$ for different temperatures $T$ and $L \times M= 60\times
120$. As can be observed, $D_{sat}$ grows with $T$ for low temperatures and
then starts to decrease when the temperature increases. This behaviour is
similar to the already discussed case with $E=50\approx \infty$, which exhibits a peak at
$T_c (E=\infty)$ (cf. \fref{fig3}). 
Then, we conclude that for a large driving field (e.g. $E\geq10$) the saturation damage becomes independent of the magnitude of the field for all the studied temperatures.
The fact that in the presence of a driving field $D_{sat}$ is smaller than the pure Ising model with Kawasaki dynamics (i.e. $E\equiv 0$) points out that the field tends to enhance the healing of damage.
This healing effect is most likely due to the existence of a preferential direction of motion, which established a macroscopic stationary current in the direction parallel to the driving.

\section{Conclusions}
\label{conclusions}

In this work we studied damage spreading in a DLG model, i.e. in an archetypical model that evolves
towards non-equilibrium stationary states. By setting the magnitude of the external field
$E=50$, which is
taken as infinite for practical purposes, we found that the damage always
propagates as a consequence of the conservative dynamics and reaches a
saturation value $D_{sat}$ that is independent of the system size and the aspect
ratio, but depends on $T$. In fact, \fref{fig3} shows that $D_{sat}<0.5$ for all temperatures, it
increases for $T<T_c$ and decreases for $T>T_c$, so it exhibits a peak close to $T_c$.
The discontinuity in the slope of $D_{sat}$ around $T_c$ 
observed for $E=\infty$ suggests that the damage is sensitive to the presence of interfaces. This fact has already been observed in equilibrium systems such as the confined magnetic systems (see
\cite{rubio2008}). In order to show this, we started the system from an ordered configuration (see \sref{results}) and studied the stationary transversal damage profiles as a function of the temperature. 
In a low temperature interval $2.5<T<2.8$ (panels (a)-(b) of \fref{fig4}), almost all the damage is concentrated in one peak located very close to the imposed interface at the middle of the lattice. 
This happens because the interface width is small (see \fref{fig5} (a)) and thus damage cannot spread all over the system. 
Then, near and at the critical temperature, $3.0\leq T\leq T_c=3.2$ (\fref{fig4} (c)-(d)), the interface width is rougher than in the previous cases (see \fref{fig5} (b)), so damage can be generated and spreads easily. 
Also, due to the diffusive nature of the system holes develop in the bulk of the high-density strips, where new interfaces can be formed and consequently create some new damaged sites.
For $T>T_c$ there are still some extended clusters in the system (\fref{fig5} (c)) with  a large scattering of their interfacial width. 
In this case the damage is no longer located at the interface of the strips but it becomes delocalized, as is shown in \fref{fig4} (e).  
Finally, at $T\gg T_c$ the temperature has reduced the size of the clusters, the interfaces are disappearing, and damage decreases (\fref{fig4} (f)).
Furthermore, the maximum value of $D_{sat}$ around $T_c$ in \fref{fig3} suggests that this observation could be employed as a new
method to obtain a rough estimate of the critical temperature in other non-equilibrium systems of
different nature.\\
Concerning damage evolution inside the high-density strips, \fref{fig6} shows that for short times its transversal motion is diffusive (i.e. from the centre to the border of the strips) and then the damage remains attached to the interface where its motion becomes essentially parallel to the driving field that, of course, establishes a macroscopic current along its preferential direction. 
In this way, we can conclude that the interfaces play a crucial role in the creation and spreading of damaged sites in the DLG model.\\ 
The behaviour described above was also observed in the range $0.5<E<50$ (see right-hand side panel of \fref{fig7}).  
In this context, damage spreading could be a useful
method to study the role of interfaces in the critical behaviour of non-equilibrium systems.
Furthermore, we have also shown that the driving field enhances the healing of damage as compared with the pure Ising model with $E=0$.
\\

\ack
This work was supported financially by CONICET, UNLP, and ANPCyT (Argentina). 

\section*{References} 

\bibliographystyle{unsrt}
\bibliography{damkls}

\end{document}